\documentclass[conference]{IEEEtran}
\IEEEoverridecommandlockouts
\usepackage{amsmath,amssymb,amsfonts}
\usepackage{algorithmic}
\usepackage{graphicx}
\usepackage{textcomp}
\usepackage{xcolor}
\usepackage[a4paper, total={184mm,239mm}]{geometry}
\def\BibTeX{{\rm B\kern-.05em{\sc i\kern-.025em b}\kern-.08em
    T\kern-.1667em\lower.7ex\hbox{E}\kern-.125emX}}
    
    \usepackage{color}
    
\usepackage{dblfloatfix}
\usepackage{textcomp}
\usepackage{gensymb}
\usepackage[subtle]{savetrees}
\usepackage[style=numeric, sortcites=true]{biblatex}
\usepackage{dirtytalk}

\usepackage[ruled,vlined]{algorithm2e}

\newcommand\nabavi[1]{\textcolor{black}{#1}} 
\newcommand\larimi[1]{\textcolor{black}{#1}} 
\newcommand\behzad[1]{\textcolor{black}{#1}} 
\newcommand\salami[1]{\textcolor{black}{#1}} 
\newcommand\seyed[1]{\textcolor{black}{#1}} 
\newcommand\sabern[1]{\textcolor{black}{#1}}
\newcommand\sabera[1]{\textcolor{black}{#1}}
 
\begin{document}

\graphicspath{./Resources}
    
\title{Understanding Power \nabavi{Consumption} and Reliability of High-Bandwidth Memory with Voltage Underscaling\\\vspace{-0.5cm}
\begin{center}
{\normalsize \hspace{2cm} Seyed Saber Nabavi Larimi\textsuperscript{1,2}\hspace{2cm} Behzad Salami\textsuperscript{1,5}\hspace{2cm} Osman S. Unsal\textsuperscript{1}\hspace{2cm}}\vspace{-0.5cm}

{\normalsize \hspace{1.8cm} Adrián Cristal Kestelman\textsuperscript{1,2,3}\hspace{1.7cm} Hamid Sarbazi-Azad\textsuperscript{4}\hspace{2cm} Onur Mutlu\textsuperscript{5}\hspace{2cm}}\vspace{-0.3cm}
    
{\normalsize \hspace{0cm}\hfill \textsuperscript{1}BSC\hspace{2cm}\textsuperscript{2}UPC \hspace{2cm}\textsuperscript{3}CSIC-IIIA\hspace{2cm} \textsuperscript{4}SUT and IPM\hspace{2cm}\textsuperscript{5}ETH Zürich\hfill\hspace{0cm}}\vspace{-1.5cm}
\end{center}
}
\author{}
\maketitle
\begin{abstract}
Modern computing devices employ High-Bandwidth Memory (HBM) to meet their memory bandwidth requirements. An HBM-enabled device consists of multiple DRAM \behzad{layers} stacked on top of one another next to \behzad{a} compute chip (e.g. CPU, GPU, and FPGA) in the same package. Although such HBM structures provide high bandwidth at a small form factor, the stacked memory \behzad{layers} consume a substantial portion of the \nabavi{package's power budget}. Therefore, power-saving techniques that preserve the performance of HBM are desirable. Undervolting is one such technique\larimi{: it} reduces the supply voltage to decrease power consumption without reducing the \nabavi{device's operating frequency} to avoid performance loss. Undervolting takes advantage of voltage guardbands put in place by manufacturers to ensure correct \nabavi{operation} under all environmental conditions. However, reducing voltage without changing frequency \behzad{can lead} to reliability issues manifested as unwanted bit flips.

In this \behzad{paper, we provide the first} experimental study of real HBM \nabavi{chips} \behzad{under} reduced-voltage conditions. We show that the guardband regions for our HBM \nabavi{chips} \behzad{constitute} 19\% of the nominal voltage. Pushing the supply voltage down \behzad{within} \larimi{the} guardband region \nabavi{reduces} power \nabavi{consumption} by a factor of 1.5X for all bandwidth utilization \behzad{rates}. Pushing \behzad{the} voltage down further by 11\% \nabavi{leads to a total of} 2.3X power saving\behzad{s} at the cost of unwanted bit flips. We explore and characterize the rate and type\behzad{s} of these \nabavi{reduced-voltage-induced} bit flips and present a \behzad{fault} map that enables the possibility of a three-factor trade-off among power, \larimi{memory} capacity, and \larimi{fault rate}.
\end{abstract}

    
\section{Introduction} \label{sec:introduction}
\behzad{Dynamic Random Access Memory (DRAM) \nabavi{is the predominant} main memory \nabavi{technology used in} traditional computing systems. \nabavi{W}ith the \larimi{significant} growth in the computational capacity of modern systems, DRAM has become a power/performance/energy bottleneck, especially for \nabavi{data-intensive} applications \nabavi{\cite{uvDRAM, IMW2013Mutlu, SUPERFRI2014Mutlu, mutlu2020modern, SIGMETRICS2019Sag}}. There are two approaches to alleviate this issue: \textbf{(i)} replacing DRAM with emerging technologies (\larimi{e.g.,} Magnetic Memory (MRAM) \nabavi{\cite{racetrack, ISPASS2013Kultursay}} and Phase-Change Memory (PCM) \nabavi{\cite{PCM, ISCA2009Lee, ISCA2009Qureshi}}) and \textbf{(ii)} improving DRAM \nabavi{design} (\larimi{e.g.,} Reduced Latency DRAM (RLDRAM) \cite{RLDRAM}, Graphics DDR (GDDR) \cite{GDDR6}, and Low-Power DDR (LPDDR) \cite{LPDDR4}). To \nabavi{the latter end}, High-Bandwidth Memory (HBM) \cite{Arxiv2016Lee, HCS2014Kim} has been developed \nabavi{to} bridge the \emph{bandwidth} gap of computing devices and DRAM-based main memory.}

An HBM-enabled device consists \nabavi{of} \behzad{multiple} DRAM \behzad{layers} \larimi{stacked} and placed next to computing elements, all integrated in the same package. Higher bandwidth, lower power consumption, and smaller form factor are the advantages of such integration. Therefore, despite being a relatively new technology, HBM has found its way into high-end devices such as NVIDIA A100 \cite{A100}, Xilinx Virtex Ultrascale+ HBM family \cite{VirtexHBM}, and AMD Radeon Pro family \cite{RadeonPro}, and \larimi{into some of the world's fastest computing systems such as the Summit supercomputer \cite{Summit}}. However, being placed inside the same package with \nabavi{computing} devices \larimi{means} that HBM \behzad{consumes} a portion of the package's overall power budget, \larimi{limiting} the power available for \larimi{computing} devices. Since \larimi{HBM targets} high-performance applications, any power saving technique with bandwidth overhead is undesirable. Therefore, \behzad{there is \nabavi{a} need for methods that save} power without \larimi{reducing} the bandwidth.

\textbf{Undervolting}, \behzad{also called voltage underscaling}, lowers supply voltage without decreasing operating frequency, \behzad{thereby} saving power without \behzad{affecting} performance. \behzad{In real devices, undervolting is effective because manufacturers conservatively specify a higher supply voltage for the operation of a device than the minimum necessary supply voltage for correct operation.} The difference between the default supply voltage and \larimi{this} minimum \larimi{supply} voltage is called \seyed{``guardband''}. \nabavi{G}uardbands are put in place to ensure \nabavi{correct and} consistent \behzad{operation under} all \nabavi{possible (including worst-case) operating} conditions. Pushing \behzad{the supply} voltage down in \larimi{the} guardband \larimi{region reduces power consumption}.

\nabavi{We} \behzad{obtain} 1.5X power savings in real HBM \behzad{chips} under all bandwidth utilization \behzad{rates} by reducing the \larimi{supply} \behzad{voltage} from the nominal 1.2V down to 0.98V\larimi{, safely} without any faults \nabavi{under common operating conditions}. Pushing \behzad{the} supply voltage \larimi{further down} to 0.85V, results in an overall 2.3X power \larimi{savings}. However, at voltages \behzad{below the} guardband \behzad{region}, device components start experiencing timing violations\larimi{, causing} unwanted bit flips. In our experiments, first bit flips occur at 0.97V. From \larimi{0.97V} to 0.84V, the number of faults increases \emph{exponentially} until almost all bits are faulty. \larimi{Between 0.84V and 0.81V}, all bits become \larimi{faulty,} while using voltages lower than 0.81V result in the failure of \behzad{entire} HBM \behzad{chips}. To save power with undervolting, we need to understand \behzad{the \seyed{occurrence rate} of faults} at each voltage level, if and how \seyed{faults} are clustered and how far we can \sabern{lower the supply voltage} \behzad{below the guardband region}.

\larimi{Undervolting} has been experimentally studied on \salami{CPUs \larimi{\cite{uvCPU, reddi2015, reddi2010, Bacha2013}}, GPUs \cite{uvGPU, LengISCA2013, Leng2015HPCA, Zou2018} and FPGAs \nabavi{\cite{uvFPGA, Salami2019, SalamiModern}},} as well as DRAMs \behzad{\cite{uvDRAM, ICAC2011David, ISCA2020Jawad, memscale2011}}\seyed{,} SRAMs \cite{uvSRAM, Yang2017}\seyed{, and NAND flash \sabern{memories \cite{caiIntelJ, cai2013threshold, cai2017error, Cai2012DATE, Cai2015DSN, Cai2017HPCA, inside2013, CaiSSDs}}}. Our work is the first experimental study of undervolting HBM \nabavi{chips}. Our main \textbf{contributions} are as follows: 

\begin{itemize}
    \item \nabavi{We} empirically measure a 19\% voltage guardband \nabavi{in} \behzad{HBM} \nabavi{chips}\larimi{. We show that undervolting within the guardband region reduces} HBM power consumption by a factor of 1.5X.
    \item \nabavi{We} \larimi{empirically examine undervolting} below the guardband \behzad{region \nabavi{in HBM chips} and} \larimi{demonstrate} a total of 2.3X power \behzad{savings} at the cost of some unwanted bit flips.
    \item We provide the first experimental fault characterization \nabavi{study} \larimi{of HBM undervolting} below \behzad{the} guardband region. We find \larimi{that \textbf{(i)}} HBM channels behave differently from \nabavi{each other with} voltage \behzad{underscaling} due to process variation\larimi{, and \textbf{(ii)}} most faults are clustered together in small regions of \behzad{HBM layers}.
    \item We provide \behzad{a fault} map that \behzad{enables} the user \behzad{to perform} a three-factor trade-off among power, fault-rate, and usable memory space. For instance, \behzad{2.3X power savings \seyed{is}} possible by sacrificing some memory space while the \behzad{remaining} memory \behzad{space} can work with \behzad{0\% to 50\% fault rate}.
\end{itemize}
    
\section{Experimental \behzad{Methodology}} \label{sec:methodology}
\subsection{Background on HBM}
\behzad{Fig. \ref{figure:hbm_overview}(a)} shows the general organization of an HBM-enabled device where several DRAM chips (and an optional IO/controller chip) are piled and interconnected by Through Silicon Vias (TSVs) \behzad{\cite{Arxiv2016Lee}}. An efficient way to utilize \behzad{an HBM stack is to place it} on a silicon interposer next to \nabavi{computing chips} (\behzad{e.g.,} FPGA, GPU, or CPU) inside the same package \behzad{\cite{JEDEC_HBM}}. Signals between HBM stack and \nabavi{computing chips} go through the \behzad{underlying} silicon interposer. As a result, there can be far more data lanes \behzad{in an HBM channel} (1024 per HBM stack\behzad{)} than a regular 64-bit DRAM \behzad{channel}, while each \behzad{HBM channel} is more efficient. \nabavi{Therefore}, HBM provides at least an order of magnitude higher bandwidth than \behzad{DDRx DRAM} \nabavi{\cite{SIGMETRICS2019Sag}} at a lower power consumption (nearly 7pJ/bit as opposed to 25pJ/bit for a \behzad{DDRx} DRAM) with a smaller form factor \cite{wp485}.

\subsection{\behzad{Testing Platform}}    
\begin{figure}[b]
    \centering
    \includegraphics[width=0.48\textwidth]{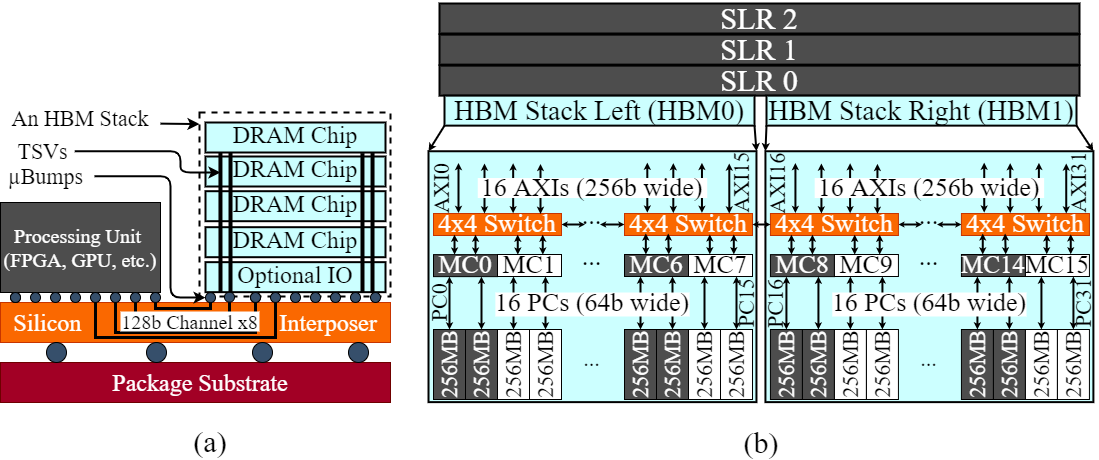}
    \vspace{-2mm}
    \caption{\behzad{(a) General structure of an HBM-enabled device. (b) HBM interface and internal organization of XCVU37P, adapted from \cite{VCU128}.}}
    \label{figure:hbm_overview}
\end{figure}

The \behzad{hardware} platform we use in our experiments consists of a Xilinx VCU128 board \cite{VCU128} mounted with an XCVU37P FPGA. This FPGA includes two HBM stacks of 4GB \behzad{each} \behzad{(HBM0 and HBM1)}. Each stack has four DRAM chips \behzad{of} 1GB capacity \behzad{each}. \behzad{Fig. \ref{figure:hbm_overview}(b)} shows a general overview of the underlying HBM memory. The FPGA fabric is divided into three Super Logic Regions (SLRs). Each SLR is a separately fabricated chip with configurable circuitry. \larimi{SLRs} are interconnected by the same \nabavi{interposer} technology connecting them to the HBM stacks. \behzad{Both HBM stacks of our setup are connected to SLR0, \nabavi{as shown in} Fig.\ref{figure:hbm_overview}(b)}.

Address space of each HBM stack is divided among 8 independent Memory Channels (MC)\nabavi{. Each MC is} 128b wide \nabavi{and works} independently on a 512MB memory assigned to \nabavi{it}. Address space of each channel \larimi{is divided} between two 64b Pseudo-Channels (PCs). These two PCs share clock and command signals but have separate data buses. \nabavi{Each PC \seyed{independently} interprets} commands and work\nabavi{s} with \nabavi{its} own non-overlapping 256MB memory array portion. Therefore, at memory side, there are a total of 32 PCs, 64b wide each. At \nabavi{user} side, Xilinx's HBM IP core provides 32 AXI ports \behzad{(16 per HBM stack)}. Each AXI port corresponds to one PC. However, if the switching network is enabled, any packet from an AXI port can be routed to any PC at the cost of extra delay and \behzad{lower} bandwidth. \larimi{An} AXI \larimi{port is 256b wide}\nabavi{,} which provides a 4:1 data width ratio over \larimi{a PC} (\behzad{with \larimi{64b} width}). As a result, \nabavi{an AXI port} can \nabavi{operate at} a clock frequency that is a quarter of \nabavi{the} memory data transfer rate \behzad{(1:4 ratio)} and yet \behzad{take} advantage of \nabavi{the} maximum HBM bandwidth provided by PCs \cite{wp485}. The maximum clock frequency allowed for memory arrays in our device is 900MHz, and being a \nabavi{double data rate} memory, it translates to a maximum data transfer rate of 1800 Mega-Transfers per second (MT/s).

\behzad{In this work}, we tune the supply voltage of our HBM stacks by accessing voltage regulators on the VCU128 board. One of these regulators, ISL68301, is a \behzad{Power Management Bus (PMBus) \cite{pmbus}} compliant driver from Intersil Corporation in charge of supplying power to our HBM stacks. We implemented a customized interface on the host to control this regulator and measure power, voltage and current during our experiments. We also implement controllers for the two HBM stacks. Each controller includes 16 AXI Traffic Generators (TG), one for each AXI port in that stack. The controller is in charge of configuring each TG, sending macro commands, receiving responses, checking status, and reporting statistics back to the host. Each TG is capable of running customized macro commands that we later use to implement our test \nabavi{routines}. We \behzad{collect} \larimi{power} measurements from \larimi{a} Texas \larimi{Instruments} INA226 chip placed on VCU128 board.

\subsection{\behzad{Experiments}}
    \label{sec:test_scenarios}
    We conduct experiments to measure \nabavi{\textbf{(i)}} the power we can save with undervolting and \nabavi{\textbf{(ii)}} \larimi{the fault rate of our HBM devices} when we reduce the voltage below the nominal value. The methodology we used for our experiments, considers the following points:
\begin{itemize}
    \item Since we \behzad{focus} on HBM stacks and not \behzad{FPGA} fabric, we \behzad{disable} the switching network. This removes any impact the switching network might \behzad{have on} the results. 
    \item \larimi{We follow a statistical method to \seyed{determine} the number of runs based on error and confidence margin \cite{Leveugle2009}. We run \seyed{each} test 130 times, which gives us a 7\% error margin with 90\% confidence interval.}
    \item \behzad{HBM bandwidth is much larger than \nabavi{the} communication speed between \nabavi{the} FPGA and  host \larimi{CPU}. As a result, we focus on measuring simple statistics on the FPGA itself and then report those raw numbers back to the host for further analysis.}
    \item The operating temperature of HBM stacks was 35 \textpm 1\degree C during \larimi{our} experiments.
\end{itemize}

\behzad{We conduct the following power and reliability experiments:}

\subsubsection{Power Measurement Tests}
    We measure the power consumption of HBM stacks at different bandwidth \behzad{utilization rates} while \behzad{underscaling} their supply voltage. \behzad{We reach} nearly 310GB/sec \nabavi{when} \behzad{accessing} the memory by enabling all 32 AXI ports at the same time and running them \behzad{at maximum frequency.}\footnote{\behzad{The combined peak theoretical bandwidth of HBM stacks in VCU128 is 429GB/sec \cite{VCU128}. For the experiments discussed in this paper, we reach the throughput of 310GB/sec\nabavi{. H}owever, we believe that with more engineering \nabavi{effort}\larimi{,} the peak performance is also achievable. The power savings obtained \nabavi{via} undervolting is achievable for any bandwidth utilization rate, as discussed in Section \ref{sec:results}.}} \behzad{We then} progressively disable \nabavi{AXI} ports to reduce bandwidth. We \behzad{do this} since some AXI ports (and their corresponding \larimi{PCs}) that map \larimi{to} the more vulnerable HBM memory blocks are more sensitive to faults induced by undervolting than others. \behzad{Therefore,} disabling those ports is an effective technique to decrease the impact of undervolting faults and further reduce the supply voltage. \nabavi{Section \ref{sec:results}\larimi{-B} discusses v}ariability \behzad{across} different ports \behzad{in} more details. 

\subsubsection{\behzad{Reliability Assessment}}
    The fault characterization \behzad{test we} conduct \nabavi{writes} data into the undervolted HBM sequentially and then \nabavi{reads} \larimi{it} back to check for any faults. Algorithm \ref{algorithm:sequential} shows the pseudo-code of \larimi{the reliability} tester \larimi{to extract $faultCount$}. We \behzad{change the HBM's supply voltage (i.e., \textit{VCC\_HBM})} \larimi{from} 1.2V \behzad{(the nominal voltage level, i.e., $V_{nom}$)} to 0.81V (minimum voltage possible for memory operation, i.e., $V_{critical}$), with 10mV step size. \behzad{We experimentally set} the \textit{batchSize}, the number of times we repeat each test to ensure consistent results, \behzad{to} 130. \nabavi{\textit{memSize}} is the size of the memory divided by 256b (\larimi{i.e., width} of \behzad{an AXI port}). \larimi{By setting \textit{dataPattern} to all 1's or all 0's, we can check for 1-to-0 or 0-to-1 bit flips, respectively.} \nabavi{\textit{dataWidth}} is \behzad{256b} since \behzad{each AXI port is} 256b wide. \behzad{Depending \nabavi{on} the type of the test, \nabavi{\textit{memSize}} takes different values, i.e., \nabavi{256M or 8M} for testing \larimi{the} entire \larimi{HBM} or \nabavi{a} single Pseudo Channel (PC), respectively.}

\begin{algorithm}
    \KwIn{batchSize: \behzad{\textit{130}}
    
    dataPattern: \behzad{\textit{all 1's \& all 0's}}
    
    dataWidth: \textit{256 (b)}
    
    memSize: \textit{256M (testing entire HBM) \& 8M (testing one PC)}
    
    }
    
    \KwOut{\larimi{$faultCount$ (at each voltage level)}}
    
    
        \For{\larimi{$voltage$} := $V_{nom}$ \nabavi{\textbf{downto}} $V_{critical}$ \seyed{in 10mV steps}}{
        VCC\_HBM := \larimi{$voltage$}\;
    \For{b := 0 \textbf{to} batchSize-1}{
        reset\_axi\_ports()\;
        \For{address := 0 \textbf{to} memSize-1}{
            writeHBM(address, dataPattern)\;
        }
        \larimi{$faultCount$} := 0\;
        \For{address := 0 \textbf{to} memSize-1}{
            data := readHBM(address)\;
            \For{i := 0 \textbf{to} dataWidth-1}{
                \If{(data[i] != dataPattern[i])}{
                \larimi{$faultCount$ += 1}\;
                }
            }
        }
        return \larimi{$faultCount$}\;
        }
    }
    \caption{\larimi{Reliability assessment via sequential access}}
    \label{algorithm:sequential}
\end{algorithm}

\section{Results} \label{sec:results}
\subsection{Power Analysis}

\behzad{\nabavi{We divide t}he total power consumption of \larimi{an HBM chip} into \nabavi{\emph{active}} and \nabavi{\emph{idle}} portions}.
\subsubsection{\behzad{Active Power}}
\textit{Active power} consumption of \behzad{a DRAM chip} is \behzad{proportional} to the square of supply voltage ($V_{dd}$)\behzad{,} as shown in \behzad{Equation} \eqref{eq:dynmaicP} \cite{MicronDDR4}. In this equation, $C_{L}$ is \behzad{the active} load capacitance\behzad{, }${f}$ is operating frequency, \behzad{and $\alpha$ is the activity factor which determines the average charge/discharge rate of the capacitor}. \behzad{Thus, with undervolting,} we expect a quadratic reduction in active power consumption. 

\begin{equation}\label{eq:dynmaicP}
    P=\behzad{\alpha\times }C_{L}\times {f}\times {V_{dd}^2} 
\end{equation}

\begin{figure}[h]
    \centering
    \includegraphics[width=0.48\textwidth]{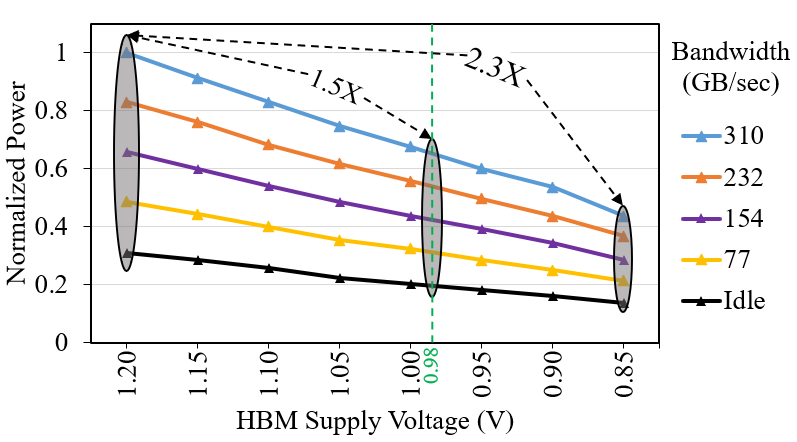}
    \vspace{-3mm}
    \caption{HBM power \behzad{saving by} undervolting. \behzad{We normalize all power measurements to the power consumption at 1.2V \nabavi{with} maximum \larimi{bandwidth utilization} (i.e., 310GB/s). \seyed{\sabern{Voltage} step size is 10mV \sabern{in our experiments, but the figure displays only the 50mV steps for better visibility.}}}}
    \label{figure:power_savings}
\end{figure}

\behzad{Our} empirical results shown in Fig. \ref{figure:power_savings} \behzad{comply with expectations}. \behzad{Fig. \ref{figure:power_savings}} shows the power consumption of HBM \nabavi{chips} at \behzad{representative} \behzad{bandwidth} utilization \behzad{rates} \behzad{(in 25\% increments).}

Working within the guardband \behzad{region} (1.20V-0.98V), provides 1.5X power savings while pushing the supply voltage further down to 0.85V results in a total of 2.3X \behzad{savings} compared to default voltage 1.2V. In both cases (within or \behzad{below the} guardband \behzad{region}), the amount of power \behzad{savings} is \behzad{\emph{independent}} of \behzad{the} bandwidth \behzad{utilization} \behzad{because undervolting does \larimi{\emph{not}} affect the \nabavi{memory} bandwidth}. Therefore, we can save the memory power by undervolting no matter what the memory bandwidth demand is.

\begin{figure}[t]
    \centering
    \includegraphics[width=0.5\textwidth]{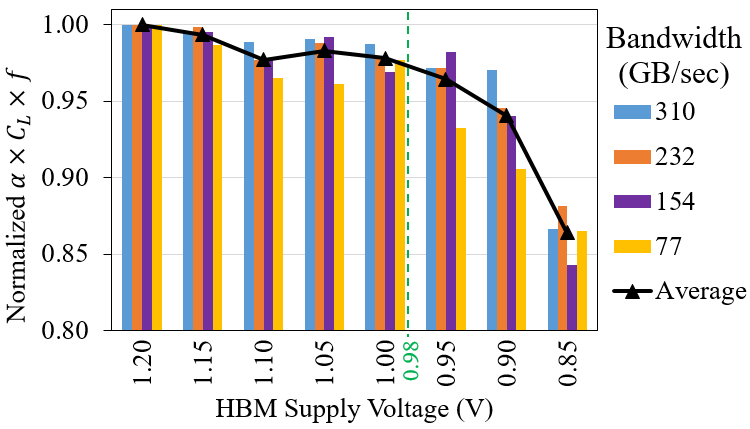}
    \vspace{-3mm}
    \caption{\behzad{Normalized \larimi{$\alpha\times C_{L}\times {f}$}. For each bandwidth, we \seyed{normalize} all values to \behzad{$\alpha\times C_{L}\times {f}$} of that bandwidth at 1.2V \behzad{to rule out the effect of bandwidth on load capacitance}. Below the guardband region, the active capacitance is lower than our expectation due to some bits \larimi{remained stuck at 0 or 1}, resulting in additional power gain.} \seyed{\sabern{Voltage} step size is 10mV \sabern{in our experiments, but the figure displays only the 50mV steps for better visibility.}}}
    \label{figure:Cxf}
\end{figure}

On the other hand, looking at \behzad{Equation} \eqref{eq:dynmaicP}, if we divide our power measurement results by $V_{dd}^2$, we are left with raw values for \behzad{$\alpha\times C_{L}\times {f}$}. The \nabavi{unit} for these values is \textit{farads per second}\behzad{,} which \behzad{indicates} how much capacitance is being \behzad{\textit{actively}} charged/discharged every second. \behzad{${f}$} is constant since the clock \behzad{frequency of} HBM memory, and the sequence that we run these tests are always fixed. \behzad{$\alpha\times C_{L}$}, on the other hand, depends on \behzad{the} memory \behzad{bandwidth utilization rate}, which we expect to remain fixed when working at a fixed bandwidth (\nabavi{i.e.,} same number of PCs). As a result, we expect values for \behzad{$\alpha\times C_{L}\times {f}$} to remain the same throughout our experiments. \behzad{However, through undervolting, we observe that HBM chips' fidelity starts to degrade. This is \nabavi{because} at voltages lower than 0.98V (i.e., below the guardband region), some bits \larimi{remain \seyed{always stuck} at 0 or 1}. Since memory operations cannot charge or discharge these faulty bits anymore, \nabavi{such bits} do \nabavi{\emph{not}} contribute to the overall active capacitance, resulting in a drop in $\alpha$. We show this behavior in Fig. \ref{figure:Cxf}}\larimi{: for} supply voltages above 0.98V, $\alpha\times C_{L}\times {f}$ \nabavi{remains} within 3\% of what we expect. \nabavi{However, below 0.98V, it starts dropping and at 0.85V it reaches 14\% lower than the maximum active capacitance (at nominal voltage)}. \behzad{In other words, undervolting below the guardband region leads to a lower active capacitance, \larimi{as shown in} Fig. \ref{figure:Cxf}. This is due to the exponential increase of fault rate in HBM memory, as discussed in \behzad{Section \ref{sub_reliability_analysis}.}}

\subsubsection{\behzad{Idle Power}}
\behzad{To evaluate} \textit{idle power} \behzad{savings}, we measure the power consumption of HBM when bandwidth utilization is zero. We find that even when HBM is idle, it consumes nearly one-third of \behzad{the power} it consumes at full load \behzad{with 100\% \nabavi{bandwidth} utilization}\behzad{, limiting the maximum amount of power we can save. As seen in Fig. \ref{figure:power_savings}, \nabavi{idle} power gradually reduces \larimi{with} undervolting}.

\begin{figure}[b]
    \centering
    \includegraphics[width=0.5\textwidth]{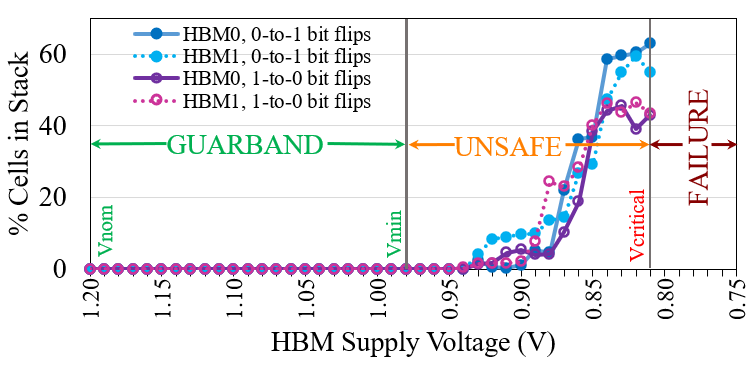}
    \vspace{-3mm}
    \caption{\behzad{Fraction of faulty portion in each HBM stack at different supply voltages.}}
    \label{figure:0vs1}
\end{figure}

\begin{figure*}
    \centering
    \includegraphics[width=0.98\textwidth]{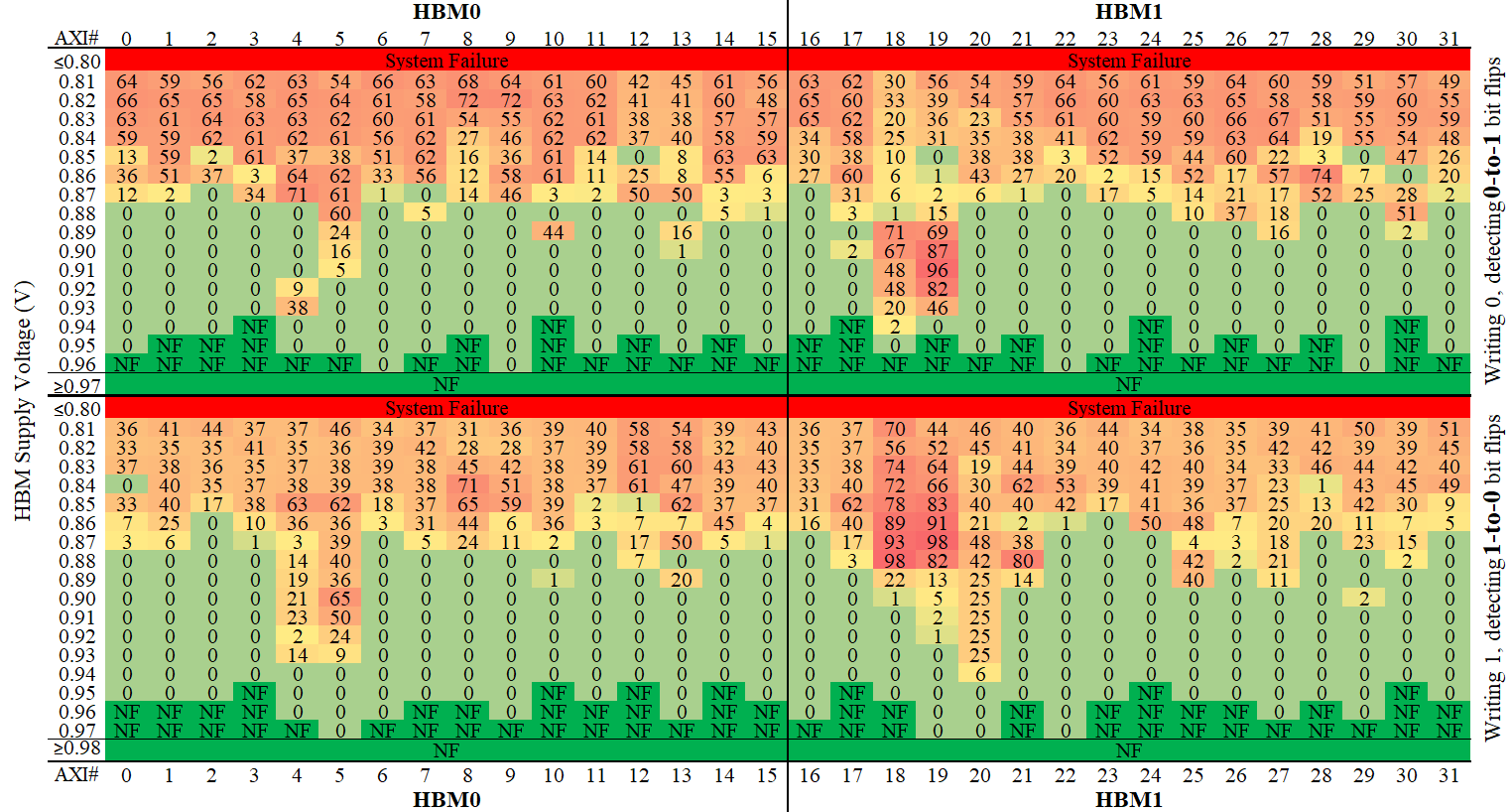}
    \vspace{-3mm}
    \caption{Percentage (\%) of memory cells that are faulty for each AXI port (and its corresponding PC) \seyed{at different supply voltages}. The left and right halves refer to HBM0 and HBM1 chips. (\seyed{Values less than 1\%} are rounded to \seyed{0\%.} \seyed{``NF''} means that \sabern{``No Fault'' is} observed.)}
    \label{figure:per_axi}
\end{figure*}

\subsection{Reliability Analysis}
\label{sub_reliability_analysis}
\subsubsection{\behzad{Overall Analysis}}
Fig. \ref{figure:0vs1} shows \larimi{the behavior of each HBM stack with undervolting}. \behzad{We observe the followings:}
\begin{itemize}
    \item \textbf{Guardband Region:} Starting from the nominal voltage ($V_{nom}$=1.2V) down to the minimum safe voltage ($V_{min}$=0.98V), \nabavi{we observe} \behzad{\emph{no}} memory faults. This \emph{guardband} region is safe for all operations \behzad{and workloads}. An application that \nabavi{cannot} tolerate any \larimi{fault in memory} has to work in this region.
    
    \item \textbf{Unsafe Region:} Faults occur in voltages below $V_{min}$. \larimi{Any application that uses HBM with supply voltages in the \emph{unsafe} region needs to take the impact of such faults into account to ensure correct operation. Reducing the voltage introduces new \larimi{faults with} an \textbf{exponentially} growing trend until about 0.84V, where all memory bits experience \larimi{0-to-1 or 1-to-0 bit flips}. Other works \seyed{have} reported similar exponential growth of faults with undervolting on regular DDRx DRAM chips \cite{uvDRAM}. \larimi{Below} 0.84V down to the minimum working voltage ($V_{critical}$=0.81V), the entire HBM parts become faulty.} 
    
    \item In our tests, HBM crashes \behzad{(i.e., stops responding)} at voltages below $V_{critical}$. Even restoring the supply voltage does not \behzad{re-enable operation}, and a power-down \behzad{and restart} is required.
\end{itemize}
\subsubsection{\larimi{Detailed Analysis \seyed{of} Fault Rate Variation}}
\larimi{Fig. \ref{figure:per_axi} shows the fault rate for each HBM chip, each AXI port (and its corresponding PC), and each data pattern at supply voltage levels below $V_{min}$. \sabern{\sabera{Due to process variation} and noise in memory, we observe three categories of }fault rate/type variation:
\begin{itemize}
    \item \textbf{Variation Across HBM Chips:} In the \emph{unsafe} voltage range (i.e., between $V_{min}$ and $V_{critical}$), \sabera{HBM0 has lower fault rate than HBM1 (13\% on average). However, both stacks have the same $V_{min}$ and $V_{critical}$.}
    \item \textbf{Variation Across PCs:} Some PCs are more sensitive to undervolting than others (e.g., PC4 and PC5 of HBM0 and PC18, PC19, and PC20 of HBM1). These PCs experience a higher rate of bit flips when we reduce the voltage below $V_{min}$.
    \item \textbf{Data Pattern Variation:} The first 1-to-0 and 0-to-1 bit flips start at 0.97V and 0.96V, respectively. \seyed{The} average rate of 0-to-1 bit flips is 21\% higher than \sabern{that of} 1-to-0 bit flips.
\end{itemize}
}


\subsection{\behzad{User- and Application-Level Implications}}
Applications that are intrinsically \behzad{resilient} to faults can save more power than others \behzad{by \nabavi{taking} advantage of aggressive} undervolting \behzad{even below the guardband region}. To \behzad{effectively achieve this \nabavi{benefit},} \nabavi{application} developers need practical information about \behzad{the effects of} undervolting. \behzad{To this end,} we present a three-factor trade-off among \textbf{power}, \textbf{fault rate}, and \textbf{available memory \behzad{capacity}} that helps application developers determine how much power can be saved and \larimi{what the associated costs are}.

\begin{figure}[b]
    \centering
    \includegraphics[width=0.5\textwidth]{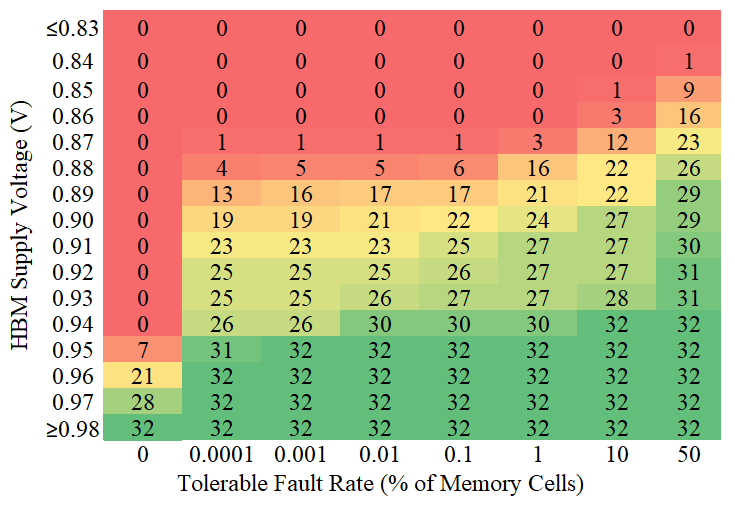}
    \vspace{-3mm}
    \caption{\behzad{Number of PCs (out of 32) that can be used under different tolerable fault \larimi{rates} with respect to the \larimi{supply} voltage level of the HBM memory. Higher numbers mean higher memory capacity and bandwidth available for applications.}}
    \label{figure:healthyPCs}
\end{figure}

\behzad{An HBM chip has multiple independently-controllable PCs} (32 in our case). We utilize this inherent independence to provide practical information about how many PCs an application can use based on its tolerable fault rate, \nabavi{as shown in} Fig. \ref{figure:healthyPCs}. \behzad{For example:}
\begin{itemize}
    \item Those applications that \behzad{cannot} tolerate any faults (e.g., \cite{Oyku2019, salamiMicpro, salamiFCCM}) and need the entire 8GB of HBM are restricted to work only in the \textit{guardband} region\nabavi{,} which starts at $V_{nom}$=1.2V and ends at $V_{min}$=0.98V. This region \behzad{offers} a fixed 1.5X power \behzad{savings} without any trade-off option. 
    
    \item \behzad{Below \nabavi{$V_{min}$},} a \seyed{triple-factor} trade-off is at the user's disposal. \seyed{For example, up to 1.6X power savings is achievable for an application that cannot tolerate \emph{any} faults but can \sabern{work with smaller} memory capacity\sabern{, by using} only 7 \sabern{fault-free} PCs \sabern{operating at} 0.95V.}
    
    \item \larimi{A}pplications \nabavi{that can tolerate a \seyed{\emph{non-zero}} fault rate (e.g., \cite{EDEN, SalamiModern, OSDI18, LouDSN2014}) allow more room for trade-offs}. For example, an application that can tolerate a 0.0001\% fault rate and requires \behzad{only} half \behzad{of the total memory capacity} can push the voltage down to 0.90V and save power by a factor of about 1.8X. 
\end{itemize}
    
\section{Related Work}\label{sec:background}
To our best knowledge, this paper presents the first experimental study of undervolting in \salami{real High-Bandwidth Memory (HBM) chips}. Below, we briefly cover closely-related work \larimi{on} reduced-voltage operation in other computing \salami{and} memory devices.

\begin{itemize}
    \item \behzad{\textbf{General-Purpose Processors:}} Papadimitriou et al. \behzad{explore} undervolting for multi-core ARM processors \cite{Papadimitriou2019, uvCPU}. They show up to 38.8\% \larimi{power reduction} at the cost of up to 25\% performance loss. \salami{Similar} undervolting studies \salami{are} conducted for other \salami{types of} \salami{processors \cite{Bertran2014, Bacha2013, reddi2009, reddi2010, reddi2015}}.
    
    \item \textbf{Hardware Accelerators:} Undervolting in FPGAs \salami{has} recently been \sabern{studied \cite{behzadIOLTS, behzadThesis, uvFPGA, SalamiECC, SalamiModern, Salami2019, Salami2019FPL}.} These studies focus on undervolting multiple components of FPGAs (e.g., Block RAMs (BRAMs) and internal \larimi{components).} \larimi{Undervolting in GPUs} is \salami{also} \larimi{studied} with detailed \seyed{analysis} of power saving, \salami{voltage \salami{guardbands}, and reliability \larimi{costs} \cite{Zou2018, Leng2015HPCA, uvGPU, LengISCA2013}.}
    
    \item \behzad{\textbf{Memory Chips:}} Koppula et al. \cite{EDEN} propose \larimi{a} \behzad{DRAM undervolting and latency reduction} framework for neural networks that \behzad{improves} energy \behzad{efficiency and performance of such networks \larimi{by} 37\% and 8\%, respectively}. Chang et al. \cite{uvDRAM} study the impact of undervolting on \behzad{the reliability and energy consumption} of DRAM by characterizing faults in \behzad{real DRAM chips} and provide techniques to mitigate \behzad{undervolting-induced faults}. \salami{Earlier} works \salami{study} undervolting \larimi{in} main memory \salami{systems \cite{ICAC2011David, memscale2011}}, but \larimi{do} not analyze faults due to undervolting. \salami{Ghose} et al. \cite{Sag2017} \larimi{study power consumption} in \salami{modern} DRAM chips. \larimi{Luo} et al. \cite{LouDSN2014} show that unreliable DRAM chips can be used in real applications to enhance \larimi{the} cost of scaling the main memory system. Undervolting is also studied for other \sabera{memory types} like SRAM \cite{uvSRAM, Yang2017} and flash \sabern{\cite{caiIntelJ, cai2013threshold, cai2017error, Cai2012DATE, Cai2015DSN, Cai2017HPCA, inside2013, CaiSSDs}}\sabera{.}
    
\end{itemize}

In addition to real chips, undervolting is studied at \larimi{the} simulation-level, e.g., for CPUs~\cite{swaminathan2017bravo,8416495}, FPGAs \cite{FPGASim}, \seyed{ASICs \cite{reagen2016minerva,zhang2018thundervolt}}, and \seyed{SRAMs \cite{Wilkerson1, Wilkerson2,AlameldeenISCA}.}
    
\section{Conclusion} \label{sec:future}
We reported the first undervolting study \salami{of} real \seyed{High-Bandwidth Memory (HBM)} chips. We demonstrated 1.5X to 2.3X power savings for such \salami{chips via} voltage underscaling below the nominal level. We measured \salami{a} voltage guardband of \salami{19\%} of the nominal voltage, and showed that eliminating it results in 1.5X power savings. We discussed that further undervolting below the guardband region provides more power savings, at the cost of unwanted bit flips in HBM cells. We explored and characterized the behavior of these bit flips (e.g., rate, type, and variation across memory channels) and presented a fault map that enables the possibility of a three-factor trade-off between power, memory capacity, and fault rate. We conclude that undervolting for \sabera{High-Bandwidth Memory} chips is very promising for future systems.

\section*{Acknowledgments}\label{sec:acknowledgments}
The research leading to these results has received funding from the European Union’s Horizon 2020 Programme under the LEGaTO Project (www.legato-project.eu), grant agreement No. 780681. \sabera{This work has received financial support, in part, from Tetramax for the LV-EmbeDL project.} \salami{This work is supported in part by funding from \larimi{SRC} and gifts from Intel, Microsoft and VMware to Onur Mutlu.}

\printbibliography

\end{document}